\documentclass[conference,twocoloum]{IEEEtran}

\usepackage{amsmath, graphics,amssymb,epsfig,subfigure}

\newtheorem{theorem}{Theorem}

\begin{document}

\title{Not Too Delayed CSIT Achieves the Optimal Degrees of Freedom}
\author{\normalsize  Namyoon Lee and Robert W. Heath Jr.\bigskip
\\
\normalsize Wireless Networking and Communications Group\\
\normalsize Department of Electrical and Computer Engineering
\\ \normalsize The University of Texas at
Austin, Austin, TX 78712 USA\\
      { \normalsize E-mail~:~namyoon.lee@utexas.edu and rheath@ece.utexas.edu} }

\maketitle
\begin{abstract}
Channel state information at the transmitter (CSIT) aids interference management in many communication systems. Due to channel state information (CSI) feedback delay and time-variation in the wireless channel, perfect CSIT is not realistic. In this paper, the CSI feedback delay-DoF gain trade-off is characterized for the multi-user vector broadcast channel. A major insight is
that it is possible to achieve the optimal degrees of freedom (DoF) gain if the delay is less than a certain fraction of the channel coherence time. This precisely characterizes the intuition that a small delay should be negligeable. To show this, a new transmission method called \emph{space-time interference alignment} is proposed, which actively exploits both the current and past CSI.
\end{abstract}

\section{Introduction}

Channel state information
at the transmitter (CSIT) is important for optimizing wireless system performance. In the
multiple-input-single-output (MISO) broadcast channel, CSIT allows the transmitter to simultaneously send multiple data symbols to
different receivers without creating mutual interference by using
interference suppression techniques \cite{Spencer}-\cite{Caire}.
Prior work on the MISO broadcast channel focused on the CSIT uncertainty caused by limited rate feedback \cite{Jindal}-\cite{Yoo} and showed there are no degrees of freedom (DoF) lost compared to the perfect CSIT case, if the CSI feedback rate per user linearly increases with signal to noise ratio (SNR) in dB scale. Meanwhile, it has been conjectured that CSIT uncertainty due to feedback delay significantly degrades the DoF gain. 

Recently, assuming only outdated CSI at the transmitter, it was shown that \cite{Maddah-Ali2} DoF gains greater than that of TDMA can be achieved in the context of MISO broadcast channel. The key idea from \cite{Maddah-Ali2} is
to exploit the perfect outdated CSIT as side-information, which allows the transmitter to align inter-user interference between the past and the currently received signals. Motivated the work in \cite{Maddah-Ali2}, extensions have been developed for other networks such as a single antenna 3-user interference channel \cite{Maleki} and multiple antenna interference channel \cite{Ghasemi}.
The common assumption of previous work \cite{Maddah-Ali2}-\cite{Ghasemi} is that the transmitter only has delayed CSI. Depending on the relative difference between CSI feedback delay and channel coherence time, however, it may be possible for the transmitter to use current CSI during a fraction of the channel coherence time as well as outdated CSI when feedback delay is less than channel coherence time. 

%
%

\begin{figure*}
\centering
\includegraphics[width=6.0in]{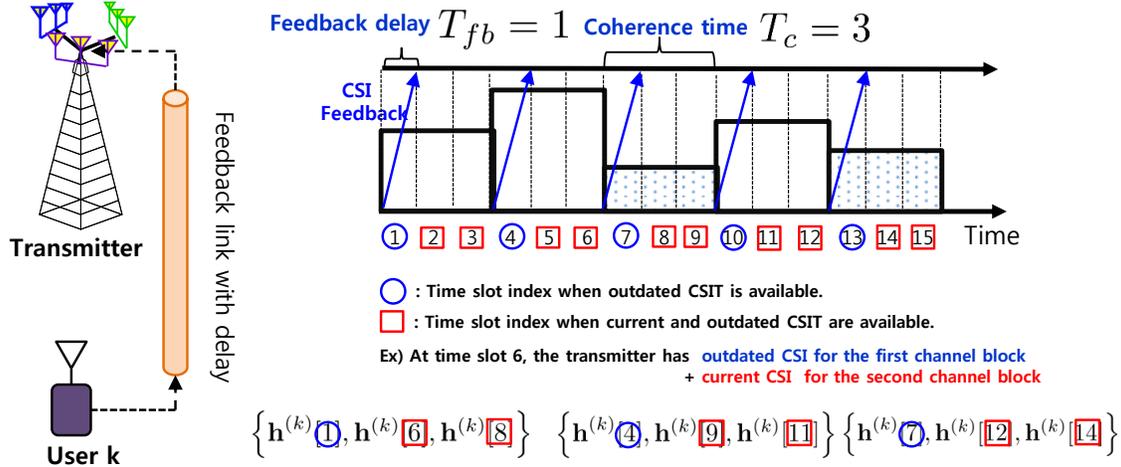}
\caption{CSI feedback model when $T_c=3$ and $T_{fb}=1$. As shown in the figure, at time slot 8, the transmitter has knowledge of current CSI for the 3-th block channel and outdated CSI for the first and second block channels.} \label{fig:system_model}
\end{figure*}


If CSI feedback is not too delayed, is it possible to increase the DoF gain by using both outdated CSI as well as current CSI? In this paper, we show that there is no DoF loss compared to the case of CSI feedback without delay, even if CSI feedback delay exists, if the delay is less than a derived fraction of the channel coherence time. For instance, we show that the $2$ of
DoF gain (cut-set outer bound) are achievable for the MISO broadcast channel
where a transmitter having $N_\textrm{t}=2$ antennas supports
$K=3$ users having a single antenna if feedback delay is less than one-third of channel coherence time. Prior work conjectured that there was always a DoF loss for any feedback delay. In our work, we show that there exists a CSI feedback delay threshold such that it does not degrade the system performance from a DoF perspective. 
 This achievability result is shown through the construction of a new transmission method called space-time interference alignment (STIA). The basic idea of STIA is to align inter-user interference signals between the past observed and the currently observed while providing linearly independent linear
combinations of the desired symbols to the corresponding users using both outdated and current CSI. Further, by using the derived result and leveraging results in \cite{Maddah-Ali2}, we characterize a CSI feedback delay-DoF gain trade-off for the vector broadcast channel. Through this trade-off analysis, we provide an insight into the interplay between CSI feedback delay and system performance from a DoF gain point of view.

%
%
%

\section{System Model}
Let us consider a $K$-user MISO broadcast channel where a transmitter with $N_\textrm{t}=K-1$ multiple antennas sends independent
messages to a receiver with a single antenna. The input-output
relationship at the $n$-th channel use is given by
\begin{eqnarray}
{y}^{(k)}[n]= {\bf h}^{(k)^T}[n]{\bf x}[n] + z^{(k)}[n],
\end{eqnarray}
where ${\bf x}[n]\in \mathbb{C}^{N_t\times 1}$ denotes the signal sent by the transmitter, ${{\bf
h}^{(k)}}[n]=\left[{h}^{(k)}_1[n], {h}^{(k)}_2[n],\ldots,{h}^{(k)}_{N_\textrm{t}}[n]\right]^T\in \mathbb{C}^{N_t\times 1}$ represents the
channel vector from the transmitter to user $k$ 
where all elements of the channel are drawn from an independent and identically
distributed (i.i.d.) continuous random variable; and $z^{(k)}[n]\sim
\mathcal{CN}(0,1)$ denotes i.i.d. Gaussian noise at user $k$ with
zero mean and unit variance for $k\in\{1,2,\ldots,K\}$. We assume that the
transmit power at the transmitter satisfies an average constraint
$\mathbb{E}\left[\textmd{Tr}\left({\bf x}[n]{{\bf
x}}^{*}[n]\right)\right] \leq P$.  Further, we assume that each mobile user has perfect CSI at the receiver.

\subsection{Block Fading Channel and CSI Feedback Delay Model}
In this paper, we assume that a block fading model where the fading channel values are constant for the channel coherence time $T_c$ and change independently between blocks. Under the block fading model, as illustrated in Fig. 1., each user feeds back CSI to the transmitter every $T_c$ time slots where $T_c$ denotes channel coherence time. If we consider feedback delay time $T_{fb} $ is less than channel coherence time, i.e., $T_{fb}<T_c$, the transmitter acquires knowledge of CSI at $T_{fb}$ time slot after the time slot sent back CSI by the users. Specifically, if a user feeds back CSI at time slot $n$, the transmitter has CSI at time slot $n+T_{fb}$ in our model.

Let us define a parameter for the ratio between the CSI feedback delay and the channel coherence time as $\gamma=\frac{T_{fb}}{T_c}.$ We refer to the case where $\gamma \geq 1$ as the completely outdated CSI regime as shown in \cite{Maddah-Ali2}. In this case, only completely outdated CSI  is available at the transmitter. We refer to the case where $\gamma=0$ as the current CSI point. Since there is no CSI feedback delay, the transmitter can employ current CSI over all time slots. As depicted in Fig. 1., if $\gamma=\frac{1}{3}$, the BS is able to exploit an instantaneous CSI over two third of channel coherence time and outdated CSI for the past channels.

\subsection{CSI Feedback Delay-DoF Trade-Off}
Since the achievable data rate of users depend on the CSI feedback delay and SNR, it can be expressed as a function of $\gamma$ and $\textrm{SNR}$. Using this notion, for codewords spanning $n$ channel uses, a rate of user $i$
$R^{(i)}(\gamma,\textrm{SNR})=\frac{\log|m^{(i)}(\gamma,\textrm{SNR})|}{n}$ is achievable if the
probability of error for the message $m^{(i)}$ approaches zero as
$n\rightarrow \infty$. The achievable rate
region $\mathcal{R}(\gamma,\textrm{SNR})$ is defined as the set of
achievable rate tuples $\mathcal{R}(\gamma,\textrm{SNR})=\left( R^{(1)}(\gamma,\textrm{SNR}),\ldots,R^{(2)}(\gamma,\textrm{SNR}),\ldots,R^{(K)}(\gamma,\textrm{SNR})\right)$. The total sum DoF characterizing the high SNR
behavior of the achievable rate region is defined as
\begin{eqnarray}
d(\gamma)=\sum_{i=1}^{K}d^{(i)}(\gamma)  =\lim_{\textrm{SNR}\rightarrow \infty}
\frac{\sum_{i=1}^{K}R^{(i)}(\gamma,\textrm{SNR})}{\log(\textrm{SNR})}.
\end{eqnarray}

\section{Space-Time Interference Alignment (STIA) using Current
 and Outdated CSI}
The purpose of this section is to present a motivating example for the special case of $K=3$ and $N_\textrm{t}=2$ to explain the idea of the
proposed algorithm, which simultaneously exploit both outdated CSI and current CSI so that each user sees the same interference pattern in space and time domains. Through this example, we prove the following theorem.\\

\begin{theorem} \label{Theorem 1}
\emph{The optimal DoF of 2 (outer bound) is achieved for the 3-user $ 2 \times 1$ vector broadcast channel if current CSI for two time slots and outdated CSI for one time slot are available at the transmitter.}
\end{theorem}

\proof In this proof, we show that 6 independent data symbols are delivered to three users over 3 time slots $\{{\bf h}^{(k)}[1],{\bf h}^{(k)}[6],{\bf h}^{(k)}[8] \}$ where the transmitter has current and outdated CSI at time slot 6 and 8 but no CSI knowledge at time slot 1 as shown in Fig. 1. Note that since the all three time slots belong to different channel coherence blocks, all elements of the channel are i.i.d. random variables.

\subsubsection{Phase One (Obtains Interference Pattern)} This phase consists of one time slot. In this starting phase, the transmitter has no knowledge of the CSI due to feedback delay. In this phase, the transmitter sends a total of six different data symbols; two of them are intended for each user. To this end, each user obtains an equation that consists of two desired symbols and four interference symbols. Consider time slot 1 as an example, the transmitter sends six independent
symbols where $s^{(1)}_1$ and
$s^{(1)}_2$ intended for user 1, $s^{(2)}_1$ and
$s^{(2)}_2$ intended for user 2, and $s^{(3)}_1$ and
$s^{(3)}_2$ intended for user 3 without preprocessing
\begin{eqnarray}
{\bf x}[1]= \sum_{k=1}^{3}{\bf s}^{(k)},
\end{eqnarray}
where ${\bf
s}^{(k)}=\left[s^{(k)}_1, s^{(k)}_2\right]^{T}$. Neglecting the noise at the receiver, each user saves the following equations
\begin{eqnarray}
y^{(1)}[1]&=& L^{(1,1)}[1] + L^{(1,2)}[1]+ L^{(1,3)}[1], \\
y^{(2)}[1]&=& L^{(2,1)}[1] + L^{(2,2)}[1]+ L^{(2,3)}[1],\\
y^{(3)}[1]&=& L^{(3,1)}[1] + L^{(3,2)}[1]+ L^{(3,3)}[1],
\end{eqnarray}
where $L^{(k,i)}[1]$ denotes a linear combination seen by user $k$ for the transmitted symbols for user $i$. Thus, the linear combinations are defined as 
\begin{eqnarray}
L^{(1,1)}[1]&=&h^{(1)}_1[1]s^{(1)}_1+h^{(1)}_2[1]s^{(1)}_2, \nonumber \\
L^{(1,2)}[1]&=&h^{(1)}_1[1]s^{(2)}_1+h^{(1)}_2[1]s^{(2)}_2,
\nonumber \\
L^{(1,3)}[1]&=&h^{(1)}_1[1]s^{(3)}_1+h^{(1)}_2[1]s^{(3)}_2,
\nonumber \\
L^{(2,1)}[1]&=&h^{(2)}_1[1]s^{(1)}_1+h^{(2)}_2[1]s^{(1)}_2,\nonumber \\
 L^{(2,2)}[1]&=&h^{(2)}_1[1]s^{(2)}_1+h^{(2)}_2[1]s^{(2)}_2,\nonumber \\
L^{(2,3)}[1]&=&h^{(2)}_1[1]s^{(3)}_1+h^{(2)}_2[1]s^{(3)}_2,\nonumber \\
L^{(3,1)}[1]&=&h^{(3)}_1[1]s^{(1)}_1+h^{(3)}_2[1]s^{(1)}_2,
\nonumber \\ 
L^{(3,2)}[1]&=&h^{(3)}_1[1]s^{(2)}_1+h^{(3)}_2[1]s^{(2)}_2,
\nonumber \\ 
L^{(3,3)}[1]&=&h^{(3)}_1[1]s^{(3)}_1+h^{(3)}_2[1]s^{(3)}_2.\nonumber 
\end{eqnarray}

\subsubsection{Phase Two (Same Interference Pattern Generation)} The second phase uses two time slots, i.e. $n\in \{6,8\}$. In this phase, the transmitter has knowledge of both current and outdated CSI thanks to feedback. Specifically, at  time slot 6 and 8, the transmitter has current CSI and outdated CSI for time slot 1.

Using this information, at time slot 6 and time slot 8, the transmitter simultaneously send two symbols for the dedicated users by using linear beamforming as
\begin{eqnarray}
{\bf x}^{(k)}[n]= \sum_{k=1}^{3}{\bf V}^{(k)}[n]{\bf s}^{(k)}, \quad n\in\{6,8\}
\end{eqnarray}
where ${\bf V}^{(k)}[n]\in \mathbb{C}^{2\times 2}$ denotes the
beamforming matrix used for carrying symbol vector ${\bf
s}^{(k)}=\left[s^{(k)}_1, s^{(k)}_2\right]^T$ at time slot $n$, where $n\in\{2,3\}$ and $k\in\{1,2,3\}$.

The main idea for designing beamforming matrix ${\bf V}^{(k)}[n]$ is to make all the receivers see the same linear combination for interference signals during time slot 1 by exploiting current and outdated CSI. For example,
user 2 and user 3 received the interference signals in the form of
$L^{(2,1)}[1]=h^{(2)}_1[1]s^{(1)}_1+h^{(2)}_2[1]s^{(1)}_2$ and
$L^{(3,1)}[1]=h^{(3)}_1[1]s^{(1)}_1+h^{(3)}_2[1]s^{(1)}_2$, which received information about user 1 at the time slot 1. Therefore, to deliver the same linear combination for the undesired symbols to user 2 and user 3 at time slot $n=6,8$, the transmitter constructs the beamforming
matrix carrying symbols, $s^{(1)}_1$ and $s^{(1)}_2$ as
\begin{eqnarray}
\left[%
\begin{array}{c}
  {{\bf h}^{(2)T}}[n] \\
  {{\bf h}^{(3)T}}[n] \\
\end{array}%
\right]{\bf V}^{(1)}[n]=\left[%
\begin{array}{c}
  {{\bf h}^{(2)T}}[1] \\
  {{\bf h}^{(3)T}}[1] \\
\end{array}%
\right].
\end{eqnarray}

\begin{figure*}
\begin{eqnarray}
\left[%
\begin{array}{c}
  {y}^{(1)}[1]-{y}^{(1)}[6]\\
  {y}^{(1)}[1]-y^{(1)}[8] \\
\end{array}%
\right]=\underbrace{\left[%
\begin{array}{c}
  {\bf h}^{(1)T}[1]-{\bf h}^{(1)T}[6]{\bf V}^{(1)}[6] \\
 {\bf h}^{(1)T}[1]-{\bf h}^{(1)T}[8]{\bf V}^{(1)}[8]\\
\end{array}%
\right]}_{{\bf H}^{(1)}_{\textrm{eff}}}\left[%
\begin{array}{c}
  s^{(1)}_1 \\
  s^{(1)}_2 \\
\end{array}%
\right]+\left[%
\begin{array}{c}
   z^{(1)}[1]-z^{(1)}[6]\\
  z^{(1)}[1]-z^{(1)}[8] \\
\end{array}%
\right].  \\\hline\nonumber
\end{eqnarray}
\end{figure*}

Similarly, to make the interfering users receive the same linear
combination of the undesired symbols, which is linearly dependent
(aligned) with the previously overheard equation during time slot 1, the beamforming matrices carrying data symbols for user 2 and user 3 are constructed to satisfy the
following space-time inter-user interference alignment conditions, which
are given by
\begin{eqnarray}
\left[%
\begin{array}{c}
  {{\bf h}^{(1)T}}[n] \\
  {{\bf h}^{(3)T}}[n] \\
\end{array}%
\right]{\bf V}^{(2)}[n]=\left[%
\begin{array}{c}
  {{\bf h}^{(1)T}}[1] \\
  {{\bf h}^{(3)T}}[1] \\
\end{array}%
\right],
\end{eqnarray}
and
\begin{eqnarray}
\left[%
\begin{array}{c}
  {{\bf h}^{(1)T}}[n] \\
  {{\bf h}^{(2)T}}[n] \\
\end{array}%
\right]{\bf V}^{(3)}[n]=\left[%
\begin{array}{c}
  {{\bf h}^{(1)T}}[1] \\
  {{\bf h}^{(2)T}}[1] \\
\end{array}%
\right].
\end{eqnarray}
Since we assume that channel coefficients are drawn from a continuous distribution, matrix inversion is guaranteed with high probability. Therefore, it is possible to construct transmit
beamforming matrices ${\bf V}^{(1)}[n]$, ${\bf V}^{(2)}[n]$ and ${\bf
V}^{(3)}[n]$ as
\begin{eqnarray}
{\bf V}^{(1)}[n]&=&\left[%
\begin{array}{c}
  {{\bf h}^{(2)T}}[n] \\
  {{\bf h}^{(3)T}}[n] \\
\end{array}%
\right]^{-1}\left[%
\begin{array}{c}
  {{\bf h}^{(2)T}}[1] \\
  {{\bf h}^{(3)T}}[1] \\
\end{array}%
\right], \\
{\bf V}^{(2)}[n]&=&\left[%
\begin{array}{c}
  {{\bf h}^{(1)T}}[n] \\
  {{\bf h}^{(3)T}}[n] \\
\end{array}%
\right]^{-1}\left[%
\begin{array}{c}
  {{\bf h}^{(1)T}}[1] \\
  {{\bf h}^{(3)T}}[1] \\
\end{array}%
\right],
\end{eqnarray}
and
\begin{eqnarray}
{\bf V}^{(3)}[n]&=&\left[%
\begin{array}{c}
  {{\bf h}^{(1)T}}[n] \\
  {{\bf h}^{(2)T}}[n] \\
\end{array}%
\right]^{-1}\left[%
\begin{array}{c}
  {{\bf h}^{(1)T}}[1] \\
  {{\bf h}^{(2)T}}[1] \\
\end{array}%
\right].
\end{eqnarray}
Therefore, if we denote ${{\bf \tilde{h}}^{(1)T}}[n]={{\bf h}^{(1)T}}[n]{\bf V}^{(1)}[n]$ and $L^{(1,1)}[n]={{\bf \tilde{h}}^{(1)T}}[n]{\bf s}^{(1)}$ for $n=6,8$, at time slot 6 and time slot 8, the received signals at user 1 are given by
\begin{eqnarray}
&&{y}^{(1)}[6]= \sum_{k=1}^{3}{{\bf h}^{(1)T}}[6]{\bf V}^{(k)}[6]{\bf s}^{(k)}\nonumber \\
&&\!\!\!\!\!\!\!\!\!=\!{{\bf h}^{(1)T}}\![6]{\bf V}^{(1)}\![6]{\bf s}^{(1)}\!\!\!+\!{{\bf h}^{(1)T}}\![6]{\bf V}^{(2)}\![6]{\bf s}^{(2)}\!\!\!+\!{{\bf h}^{(1)T}}\![6]{\bf V}^{(3)}\![6]{\bf s}^{(3)} \nonumber \\
&&\!\!\!\!\!\!\!\!\!={{\bf \tilde{h}}^{(1)T}}[6]{\bf s}^{(1)}+{{\bf {h}}^{(1)T}}[1]{\bf s}^{(2)}+{{\bf {h}}^{(1)T}}[1]{\bf s}^{(3)}  \nonumber \\
&&\!\!\!\!\!\!\!\!\!=L^{(1,1)}[6] + L^{(1,2)}[1]+ L^{(1,3)}[1],\\
&&{y}^{(1)}[8]= \sum_{k=1}^{3}{{\bf h}^{(1)T}}[8]{\bf V}^{(k)}[8]{\bf s}^{(k)}\nonumber \\
&&\!\!\!\!\!\!\!\!\!=\!{{\bf h}^{(1)T}}\![8]{\bf V}^{(1)}\![8]{\bf s}^{(1)}\!\!+\!{{\bf h}^{(1)T}}\![8]{\bf V}^{(2)}\![8]{\bf s}^{(2)}\!\!+\!{{\bf h}^{(1)T}}\![8]{\bf V}^{(3)}\![8]{\bf s}^{(3)} \nonumber \\
&&\!\!\!\!\!\!\!\!\!={{\bf \tilde{h}}^{(1)T}}[8]{\bf s}^{(1)}+{{\bf {h}}^{(1)T}}[1]{\bf s}^{(2)}+{{\bf {h}}^{(1)T}}[1]{\bf s}^{(3)}  \nonumber \\
&&\!\!\!\!\!\!\!\!\!=L^{(1,1)}[8] + L^{(1,2)}[1]+ L^{(1,3)}[1].
\end{eqnarray}

If we denote ${{\bf \tilde{h}}^{(2)T}}[n]={{\bf h}^{(2)T}}[n]{\bf V}^{(2)}[n]$ and $L^{(2,2)}[n]={{\bf \tilde{h}}^{(2)T}}[n]{\bf s}^{(2)}$ for $n=6, 8$, the received signals at user 2 during time slot 6 and 8 are given by
\begin{eqnarray}
&&{y}^{(2)}[6]= \sum_{k=1}^{3}{{\bf h}^{(2)T}}[6]{\bf V}^{(k)}[6]{\bf s}^{(k)}\nonumber \\
&&\!\!\!\!\!\!\!\!\!=\!{{\bf h}^{(2)T}}\![6]{\bf V}^{(1)}\![8]{\bf s}^{(1)}\!\!+\!{{\bf h}^{(2)T}}\![6]{\bf V}^{(2)}\![8]{\bf s}^{(2)}\!\!+\!{{\bf h}^{(2)T}}\![6]{\bf V}^{(3)}\![6]{\bf s}^{(3)} \nonumber \\
&&\!\!\!\!\!\!\!\!\!={{\bf {h}}^{(2)T}}[1]{\bf s}^{(1)}+{{\bf {\tilde h}}^{(2)}}[6]{\bf s}^{(2)}+{{\bf {h}}^{(2)T}}[1]{\bf s}^{(3)}  \nonumber \\
&&\!\!\!\!\!\!\!\!\!=L^{(2,1)}[1] + L^{(2,2)}[6]+ L^{(2,3)}[1],\\
&&{y}^{(2)}[8]= \sum_{k=1}^{3}{{\bf h}^{(2)T}}[8]{\bf V}^{(k)}[8]{\bf s}^{(k)}\nonumber \\
&&\!\!\!\!\!\!\!\!\!=\!{{\bf h}^{(2)T}}\![8]{\bf V}^{(1)}\![8]{\bf s}^{(1)}\!\!+\!{{\bf h}^{(2)T}}\![8]{\bf V}^{(2)}\![8]{\bf s}^{(2)}\!\!+\!{{\bf h}^{(2)T}}\![8]{\bf V}^{(3)}\![8]{\bf s}^{(3)} \nonumber \\
&&\!\!\!\!\!\!\!\!\!={{\bf {h}}^{(2)T}}[1]{\bf s}^{(1)}+{{\bf {\tilde h}}^{(2)T}}[8]{\bf s}^{(2)}+{{\bf {h}}^{(2)T}}[1]{\bf s}^{(3)}  \nonumber \\
&&\!\!\!\!\!\!\!\!\!=L^{(2,1)}[1] + L^{(2,2)}[8]+ L^{(2,3)}[1].
\end{eqnarray}

Finally, for user 3, if we denote ${{\bf \tilde{h}}^{(3)T}}[n]={{\bf h}^{(3)T}}[n]{\bf V}^{(3)}[n]$ and $L^{(3,3)}[n]={{\bf \tilde{h}}^{(3)T}}[n]{\bf s}^{(3)}$ for $n=6,8$, the received signals at time slot 6 and 8 are given by
\begin{eqnarray}
&&{y}^{(3)}[6]= \sum_{k=1}^{3}{{\bf h}^{(3)T}}[6]{\bf V}^{(k)}[6]{\bf s}^{(k)}\nonumber \\
&&\!\!\!\!\!\!\!\!\!=\!{{\bf h}^{(3)T}}\![6]{\bf V}^{(1)}\![8]{\bf s}^{(1)}\!\!+\!{{\bf h}^{(2)T}}\![8]{\bf V}^{(2)}\![8]{\bf s}^{(2)}\!\!+\!{{\bf h}^{(3)T}}\![6]{\bf V}^{(3)}\![6]{\bf s}^{(3)} \nonumber \\
&&\!\!\!\!\!\!\!\!\!={{\bf {h}}^{(3)T}}[1]{\bf s}^{(1)}+{{\bf { h}}^{(3)T}}[1]{\bf s}^{(2)}+{{\bf {\tilde h}}^{(3)T}}[6]{\bf s}^{(3)}  \nonumber \\
&&\!\!\!\!\!\!\!\!\!=L^{(3,1)}[1] + L^{(3,2)}[1]+ L^{(3,3)}[6],
\end{eqnarray}
\begin{eqnarray}
&&{y}^{(3)}[8]= \sum_{k=1}^{3}{{\bf h}^{(3)T}}[8]{\bf V}^{(k)}[8]{\bf s}^{(k)}\nonumber \\
&&\!\!\!\!\!\!\!\!\!=\!{{\bf h}^{(3)T}}\![8]{\bf V}^{(1)}\![8]{\bf s}^{(1)}\!\!+\!{{\bf h}^{(3)T}}\![8]{\bf V}^{(2)}\![8]{\bf s}^{(2)}\!\!+\!{{\bf h}^{(3)T}}\![8]{\bf V}^{(3)}\![8]{\bf s}^{(3)} \nonumber \\
&&\!\!\!\!\!\!\!\!\!={{\bf {h}}^{(3)T}}[1]{\bf s}^{(1)}+{{\bf {\tilde h}}^{(3)T}}[1]{\bf s}^{(2)}+{{\bf {\tilde h}}^{(3)T}}[8]{\bf s}^{(3)}  \nonumber \\
&&\!\!\!\!\!\!\!\!\!=L^{(3,1)}[1] + L^{(3,2)}[1]+ L^{(3,3)}[8].
\end{eqnarray}

\subsubsection{Decoding}
Now, let us consider decoding at user 1.  User 1 already has knowledge of the
interference signal $L^{(1,2)}[1]$ and $L^{(1,3)}[1]$ acquired from time slot 1. From the phase 2, user 1 received the same interference signals $L^{(1,2)}[1]$ and $L^{(1,3)}[1]$ at time slot 2 and 3 as shown in (15) and (16). Therefore, to decode the desired signal, interference cancellation is performed as
\begin{eqnarray}
y^{(1)}[1]\!-\!y^{(1)}[6]&\!\!\!\!=\!\!\!\!&L^{(1,1)}[1] + L^{(1,2)}[1]+ L^{(1,3)}[1] \nonumber \\
&-&\!\!\!\!\! L^{(1,1)}[6] - L^{(1,2)}[1] - L^{(1,3)}[1] \nonumber \\
&=&\!\!\!\!\! L^{(1,1)}[1] - L^{(1,1)}[6]
\nonumber \\
&=&\!\!\!\!\! \left({\bf h}^{(1)T}[1]-{\bf h}^{(1)T}[6]{\bf V}^{(1)}[6]\right){\bf s}^{(1)},
\end{eqnarray}
\begin{eqnarray}
y^{(1)}[1]\!-\!y^{(1)}[8]&\!\!\!\!=\!\!\!\!&L^{(1,1)}[1] + L^{(1,2)}[1]+ L^{(1,3)}[1] \nonumber \\
&-&\!\!\!\!\! L^{(1,1)}[8] - L^{(1,2)}[1] - L^{(1,3)}[1] \nonumber \\
&=&\!\!\!\!\! L^{(1,1)}[1] - L^{(1,1)}[8]
\nonumber \\
&=&\!\!\!\!\! \left({\bf h}^{(1)T}[1]-{\bf h}^{(1)T}[8]{\bf V}^{(1)}[8]\right){\bf s}^{(1)}.
\end{eqnarray}
After removing the interference signals, the effective
channel input-output relationship for user 1 during the three time slots is given in (9) (Please see the top of the this page). Since beamforming matrix ${\bf V}^{(1)}[n]$ for $n=6,8$ was designed regardless of the current direct channel ${\bf h}^{(1)T}[1]$, the elements of the effective channel vector observed at the time slot 6 and 8, i.e., $\left[\tilde{h}^{(1)}_1[6],
\tilde{h}^{(1)}_2[6]\right]={{\bf h}^{(1)T}}[6]{\bf V}^{(1)}[6]$ and $\left[\tilde{h}^{(1)}_1[8],
\tilde{h}^{(1)}_2[8]\right]={{\bf h}^{(1)T}}[8]{\bf V}^{(1)}[8]$ are also statistically independent random variables. This implies that the three channel vectors, ${\bf h}^{(1)T}[1]$, ${{\bf h}^{(1)T}}[6]{\bf V}^{(1)}[6]$, and ${{\bf h}^{(1)T}}[8]{\bf V}^{(1)}[8]$ are linearly independent. Therefore, 
$\textrm{rank}\left({\bf H}^{(1)}_{\textrm{eff}}\right)=2$ with probability one. As a result, user 1 decodes two desired symbols within three time
slots. In the same way, user 2 and user 3 are able to retrieve a
linear combination of their desired symbols by removing the
interference signals and can use the same decoding method. Since the transmitter has delivered two
independent symbols for its intended user in three channel uses,
a total $d=\frac{6}{3}=2$ DoF are achieved.\endproof

Now we make several remarks about the STIA algorithm.
\textbf{Remark 1 (Role of outdated CSI)}: The role of
outdated CSI is to provide opportunity to exploit the overheard
interference signals as side information. Specifically, by using
not only current but also outdated CSI, the transmitter can
construct the beamforming matrix for STIA so that the currently sending interference signals should be the same with the previously seen interference signals. Therefore, the received interference signals during the second phase can be perfectly eliminated from the saved interference equation in the first phase as side information. This leads to an increase in the DoF  due to exploitation of the delayed CSI feedback.


\textbf{Remark 2 (Comparison with MAT method in \cite{Maddah-Ali2})}:
Due to the requirement for current CSI, our CSI assumption is more restrictive than that demanded in \cite{Maddah-Ali2}. The proposed algorithm, however, reduces  additional CSI feedforward overhead in \cite{Maddah-Ali2}: it does not need to swap the linear
combinations of the desired symbols to obtain a new observation of the desired symbols.

\textbf{Remark 3 (Comparison with transmission algorithms using imperfect current and outdated CSI in \cite{Yang} and \cite{Gou_outdated})}:
New transmission methods combining MAT in \cite{Maddah-Ali2} and ZF method using both current and outdated CSI were developed for the two-user vector broadcast channel in \cite{Yang} and \cite{Gou_outdated}. Main difference with our assumption is that imperfect current CSI estimated by using temporal channel correlations is used in transmission algorithms \cite{Yang} and \cite{Gou_outdated}. Meanwhile, our transmission algorithm exploits perfect current CSI. Because of different channel knowledge assumption about current CSI, the algorithms in \cite{Yang} and \cite{Gou_outdated}  cannot achieve the optimal DoF for the vector broadcast channel when CSI feedback delay exists.

\textbf{Remark 4 (Connection with index coding problem)}: The index coding problem was introduced in \cite{Birk} and has been studied in subsequential work \cite{Yossef}. Further, the  index coding was studied from network coding  \cite{Rouayheb} and interference alignment \cite{Maleki2} point of view, respectively. An index coding problem is a follows: when a transmitter has a set of information messages $W=\{W_1,W_2,\ldots,W_K\}$ for multiple receivers and each receiver wishes to receive a subset of $W$ while knowing some another subset of $W$ as side information. The index coding problem is to design the best encoding strategy at the transmitter, which minimizes the minimum number of transmissions while ensuring that all receivers can obtain the desired messages. The proposed algorithm has the same objective with index coding algorithms developed in \cite{Birk}-\cite{Yossef}. This is because during phase one, each user acquires side information as form of linear combination of all transmitted data symbols where the linear coefficients are created by wireless channel. The main transmission algorithm during the second phase is to minimize the number of transmissions while ensuring that each user resolves the desired data symbols by using outdated and current CSI. Here, the system can minimize the number of transmissions during the second phase by using the beamforming that converts current channel into outdated channel going through interference symbols, which allows that each user eliminates interference signals observed during the second phase based on side information acquired at the first phase.

\textbf{Remark 5 (Implementation Issue)}
Since in the DoF analysis it is assumed that the transmitter sends the signal with large enough power, the beamforming solutions containing matrix inversion do not violate the transmit power constraint. In practice, however, when the transmitter has a finite power constraint, we need to modify the proposed algorithm so that the power constraint is satisfied. This modified algorithm may occur the performance loss but does not affect to the DoF gain. 

\section{A CSI Feedback Delay-DoF Gain Trade-Off}\label{sec:tradeoff}

In this section, we characterize a CSI feedback delay and DoF gain trade-off for the 3-user $2\times 1$ MISO broadcast channel by using the proposed algorithm in the previous section. We first prove the following theorem.
\begin{theorem} \label{Theorem 2}
\emph{There is no DoF loss for the 3-user $2\times 1$ MISO broadcast channel when $T_{fd}\leq \frac{T_{c}}{3}$, i.e.,
\begin{eqnarray}
d(\gamma) = 2, \quad {\textrm for} \quad 0<\gamma \leq \frac{1}{3}.
\end{eqnarray}}
\end{theorem}

\proof
Recall that zero-forcing (ZF) achieves the optimal DoF when CSI feedback delay does not exist, i.e., $d(0)=2$. Therefore, if we can show that $d(\frac{1}{3})=2$ by using the proposed algorithm, it is possible to show that $d(\frac{1}{3})=2$ for the region of $ 0<\gamma \leq \frac{1}{3}$ by using time sharing between the proposed STIA and ZF. Therefore, we only need to show whether $d(\frac{1}{3})=2$.
Without loss of generality, in this proof, we assume that the duration of the channel coherence is three time slots $T_c=3$ and the feedback delay time is one time slot $T_{fb}=1$, i.e., $\gamma=\frac{1}{3}$. Due to one time slot feedback delay, the transmitter can acquire CSI at $n+1$ and $n+2$ time slots if the user sends back CSI at time slot $n$. Under this channel knowledge assumption, we show that $d(\frac{1}{3})=\frac{6}{3}=2$ of DoF are achievable. The key idea is to divide total time slots into different
subsets of slots. According to different subsets of time
slots, we apply different transmission strategies: the proposed STIA, ZF and TDMA.
%

\textbf{Time Resources for STIA:} For the STIA algorithm, 
the transmitter can send two independent data symbols per user by spending three time slots where one is outdated
CSI and two are current CSI at the transmitter. Therefore, we  count a set of time slots where one is outdated
CSI and two are current CSI at the transmitter for applying the STIA. Suppose that the total number of time slots is $3n+6$ where $n$ is a large positive integer. Let us define an index set
$I_k=\{3k-2,3k+3,3k+5\}$ where $k \in
\{1,2,\ldots,n\}$. For example, as shown in Fig. 1., if
$n=3$, there exists total $15$ time resources and three index sets can be defined as $I_1=\{1,6,8\}$,
$I_2=\{4,9,11\}$, and $I_3=\{7,12,14\}$, respectively.
According to the definition of the index set, the first element, $3k-2$, corresponds to the case when outdated CSI is available at the transmitter, while the second two elements, $3k+3$ and $3k+5$, corresponds to the case when current CSI and outdated CSI is available at the transmitter. Thus, we apply the proposed STIA by using the time index sets. Notice that since we assume that the total available time resource
is $3n+6$, $3n$ time slots, i.e. $|\{I_1 \cup I_2, \cup \ldots, \cup I_{n}\}|=3n$ exist
for applying STIA. As a result, it
is possible for the transmitter to deliver $6n$ independent
symbols to three users by spending $3n$ time slots among the total $3n+6$ time slots.

\textbf{Time Resources for ZF and TDMA:} 
Since the $3n$ time slots have been used for STIA among the total time resource $3n+6$,
the remaining time resources become $3n+6-3n=6$ time
slots. Let us express the remaining time slots in terms of index as
\begin{eqnarray}
I_{R}&=&\{1,2,\ldots, 3n+6\}-\{I_1 \cup I_2, \cup \ldots, \cup I_{n}\} \nonumber \\
&=&\{2,3,5,3n+1, 3n+4,3n+6\}.
\end{eqnarray}
Recall that for the time index of $3k+1$ where $k$ is a positive integer, the transmitter sends data by using TDMA because CSIT is not available due to feedback delay. Alternatively, the transmitter delivers multiple data streams by using ZF because the transmitter is able to use CSI during the other time slots excepting the $3k+1$-th time slot. Using this observation, we decompose the remaining index set $I_{R}$ into two index sets for ZF and TDMA transmission as
\begin{eqnarray}
I_{R}=I_{ZF}\cup I_{TDMA},
\end{eqnarray}
where $I_{ZF}=\{2,3,5,3n+6\}$ and $I_{TDMA}=\{3n+1,3n+4\}$.
For the time slots in $I_{ZF}$, the transmitter sends two
independent data symbols by using  ZF beamforming. Therefore, it is possible to send the
a total $8$ data symbols by spending four slots. For the time slots in $I_{TDMA}$, the transmitter sends one data stream to one user.
\begin{figure}
\centering
\includegraphics[width=3.4in]{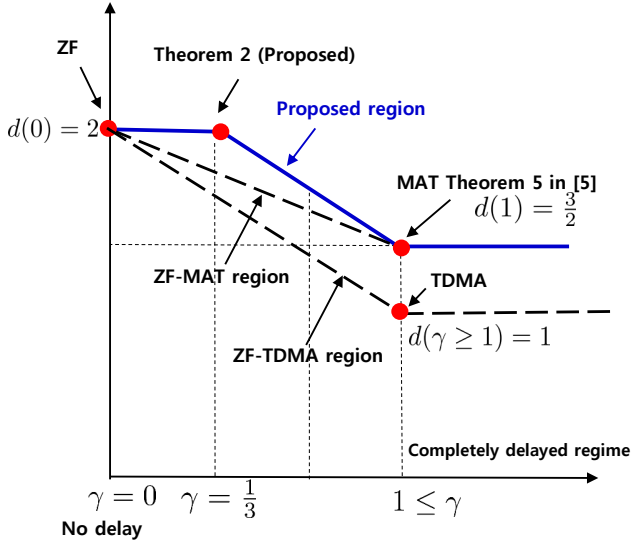}
\caption{CSI feedback delay-DoF gain trade-off for the 3-user
$2\times 1$ MISO broadcast channel.} \label{fig:Trade-off_3user}
\end{figure}

\textbf{Asymptotic the DoF gain:}
We have divided the total time resource $3n+6$ into three different groups: $N_{\textrm{STIA}}=3n$,
$N_{\textrm{ZF}}=4$, and $N_{TDMA}=2$ according to the
different transmission methods that apply. Hence, using the time sharing, the total DoF gain can be achieved by spending the $3n+6$ time
resources is given by
\begin{eqnarray}
d\left(\frac{1}{3}\right)=\frac{\underbrace{\frac{6}{3}\times 3n}_{\textrm{STIA}}+\underbrace{2\times 4}_{\textrm{ZF}}+\underbrace{1\times 2}_{\textrm{TDMA}}  }{3n+6}=\frac{6n+10}{3n+6}.
\end{eqnarray}
Therefore, as $n$ goes to infinity, the system achieves $d(\frac{1}{3})=2$ of DoF gain asymptotically.
\endproof

We interpret the result in
Theorem 2 by characterizing a CSI feedback delay-DoF gain trade-off for a
three user $2\times 1$ MISO broadcast channel.

\begin{theorem} \label{Theorem 3}
\emph{A CSI feedback delay-DoF gain trade-off for the 3-user $2\times 1$ MISO broadcast channel is given by
\begin{eqnarray}
d(\gamma) = \left\{
\begin{array}{l l}
  2, & \quad \textrm{for} \quad 0\leq \gamma\leq \frac{1}{3}, \\
  -\frac{3}{4}\gamma+\frac{9}{4}, & \quad \textrm{for} \quad \frac{1}{3}< \gamma\leq 1,\\
  \frac{3}{2}, & \quad \textrm{for} \quad \gamma\geq 1.\\
\end{array} \right.\\ \nonumber
\end{eqnarray}}
\end{theorem}

\proof From Theorem 2, $d(\frac{1}{3})=2$ of DoF are achievable when the CSI feedback delay is one-third of the channel coherence time. Further, when CSI feedback delay does
not exists, i.e., $\gamma=0$, $d(0)=2$ of DoF gain are  achieved by a conventional ZF beamforming when $N_\textrm{t}=2$ and $K=3$. When $\gamma\geq 1$, (completely
outdated delay regime), $d(\gamma\geq 1)=\frac{3}{2}$ of DoF
gain is achievable by the transmission method proposed in [\cite{Maddah-Ali2} Theorem 5 ].  It is possible to achieve any points in the line connecting three points between $d(0)$, $d(\frac{1}{3})$, and $d(1)$ by using time-sharing. The result is illustrated in Fig. \ref{fig:Trade-off_3user}.\endproof

\textbf{Remark 6 (Comparison with other algorithms)}: Let us consider a conventional transmission method, which uses ZF when current CSIT is available (time slots with blue circle in Fig. 1.) and TDMA when current CSI is unknown to the transmitter (time slots with red square in Fig. 1.). By time sharing between ZF and TDMA, it is possible to show that the $d_{ZF-TDMA}(\frac{1}{3})=\frac{2\times 2 +1\times {1}}{3}=\frac{5}{3}$ of DoF are achievable when $\gamma=\frac{1}{3}$. Similarly, if we consider time sharing method between ZF and MAT, it is possible to show that the $d_{ZF-MAT}(\frac{1}{3})=\frac{2\times 2 + \frac{3}{2}\times {1}}{3}=\frac{11}{6}$ of DoF are achieved when $\gamma=\frac{1}{3}$. Since the proposed algorithm achieves the 2 of DoF when $\gamma=\frac{1}{3}$, we obtain the $\frac{1}{3}$ of DoF gain over ZF-TDMA and $\frac{1}{6}$ of DoF gain over ZF-MAT, respectively. Therefore, as illustrated in Fig. \ref{fig:Trade-off_3user}, the proposed transmission algorithm achieves the higher CSI feedback delay-DoF trade-off region than that obtained by the other transmission techniques.

\textbf{Remark 7}: The proposed CSI feedback delay-DoF gain trade-off shows that if users feedback CSI to the transmitter within 33$\%$ of channel coherence time, the system performance is not degraded from a DoF perspective. 

\textbf{Example}: If we consider a LTE system using $f=2.1$ GHz carrier frequency, which serves users with mobility of $v=3$ km/h (walking speed). In this case the channel coherence time can be roughly calculated as $T_c \simeq \frac{c}{8fv}= 21.4$ msec (two radio frames) where $c$ denotes the speed of light. Therefore, if the users can feedback CSI within 7.133 msec (7 subframes), the performance loss does not occur from a DoF point of view. From this observation, the proposed STIA algorithm can be interpreted as a CSI delay robust transmission algorithm.


\section{Generalization of STIA}
In this section, we generalize the STIA algorithm for $K>3$ and $N_{t}=K-1$. For the case of multiple receive antennas, the similar generalization of STIA is studied in \cite{Namyoon}.

\begin{theorem} \label{Theorem 4}
\emph{$\min\{K,N_t\}=K-1$ DoF are achieved for the $K$-user $(K-1) \times 1$ vector broadcast channel if current CSI for $K-1$ time slots and oudated CSI for one time slot are avaiable at the transmitter.}
\end{theorem}

\subsection{Proof of Theorem 4}The proof is shown by the SITA algorithm. Here, we provide the proof by interpreting the proposed STIA into an index coding method.

\subsubsection{Phase One (Provide Side-Information to All Receivers)}
During phase one, the transmitter sends $K(K-1)$ independent messages to all $K$ users, $K-1$ of them are intended for each user. The main goal of this phase is to provide side-information to all users in the form of superposition of all transmitted data symbols. The transmitted signal during time slot 1 is given by
\begin{eqnarray}
{\bf x}[1]= \sum_{k=1}^{K}{\bf s}^{(k)},
\end{eqnarray}
where ${\bf
s}^{(k)}=\left[s^{(k)}_1, s^{(k)}_2, \ldots, s^{(k)}_{K-1}\right]^{T}$.
Thus, the received signal at user $k$ in time slot 1 is given by
\begin{eqnarray}
y^{(k)}[1]&=& {\bf h}^{(k)T}[1]\sum_{k=1}^{K}{\bf s}^{(k)}, \quad k\in\{1,2,\ldots, K\}\nonumber \\
&=&L^{(k,k)}[1]+\sum_{i=1,i\neq k}^{K}L^{(k,i)}[1],
\end{eqnarray}
where $L^{(k,i)}[1]={\bf h}^{(k)T}[1]{\bf s}^{(i)}$ denotes the linear combination received at user $k$ corresponding to user $i$'s signal.

\subsubsection{Phase Two (Minimize the Number of Transmissions)} In this phase, the objective is to minimize the number of transmissions by using the fact that all receivers have side-information after phase one. Recall that if we use TDMA transmission during the phase two, the required number of transmissions are $K(K-1)$ time slots because a total $K$ users want to obtain $K-1$ data streams. By using side-information obtained in the phase one and current CSI during the second phase, however, our transmission algorithm reduces the required number of transmissions as $K-1$ time slots during the second phase. The key reducing the number of transmissions is that the transmitter generates the transmit signal during the second phase so that each user sees the same pattern of interference observed at time slot 1. To accomplish this, the transmit beamforming for carrying the data symbols for user $k$ at time slot $n$, is constructed as
\begin{eqnarray}
{\bf V}^{(k)}[n]&=&\left[%
\begin{array}{c}
  {{\bf h}^{(1)T}}[n] \\
\vdots \\
  {{\bf h}^{(k-1)T}}[n] \\
  {{\bf h}^{(k+1)T}}[n] \\
  \vdots\\
  {{\bf h}^{(K)T}}[n] \\
\end{array}%
\right]^{-1}\left[%
\begin{array}{c}
  {{\bf h}^{(1)T}}[1] \\
\vdots \\
  {{\bf h}^{(k-1)T}}[1] \\
  {{\bf h}^{(k+1)T}}[1] \\
  \vdots\\
  {{\bf h}^{(K)T}}[1] \\
\end{array}%
\right],
\end{eqnarray}
where $k\in\{1,2\ldots,K\}$ and $n\in\{2,\ldots,K\}$. As shown in (30), the proposed beamforming solution converts current channel response at time slot $n$ into the past channel response at time slot 1.
From this beamforming, each user sees the same interference pattern during the second phase with the received interference pattern at time slot 1. The received signal at user $k$ at time slot $n$ is given by
\begin{eqnarray}
y^{(k)}[n]&=& {\bf h}^{(k)T}[n]\sum_{k=1}^{K}{\bf V}^{(k)}[n]{\bf s}^{(k)}, \nonumber \\
&=&L^{(k,k)}[n]+\sum_{i=1,i\neq k}^{K}L^{(k,i)}[n],\nonumber \\
&=&L^{(k,k)}[n]+\sum_{i=1,i\neq k}^{K}L^{(k,i)}[1].
\end{eqnarray}
Recall that during the second phase, i.e., $n\in\{2,3,\ldots,K\}$, user $k$ sees the same shape of interference $\sum_{i=1,i\neq k}^{K}L^{(k,i)}[1]$, which was previously saved side information at time slot 1.

\subsubsection{Decoding (Interference Cancellation)} Since each receiver has seen the same interference signal during both phases, i.e., $K$ time slots, each user is able to retrieve the desired equations by using interference cancellation technique. By using the saved equation at time slot 1, each user subtracts the interference equations received during the second phase. For example user $k$ obtains an desired equation from $y^{(k)}[2]$ by using side information acquired during the first phase $y^{(k)}[1]$ as
\begin{eqnarray}
 y^{(k)}[2]- y^{(k)}[1] &=&L^{(k,k)}[2]+\sum_{i=1,i\neq k}^{K}L^{(k,i)}[1], \nonumber \\
&-&L^{(k,k)}[1]-\sum_{i=1,i\neq k}^{K}L^{(k,i)}[1],
\nonumber \\
&=&L^{(k,k)}[2]-L^{(k,k)}[1].
\end{eqnarray}
By applying this interference cancellation for all observations, user $k$ has the following $K-1$ equations, i.e.,
 \begin{eqnarray}
\!\!\!\!\!\left[%
\begin{array}{c}
 y^{(k)}[2]- y^{(k)}[1]\\
 y^{(k)}[3]- y^{(k)}[1]\\
\vdots \\
 y^{(k)}[K]- y^{(k)}[1]\\
\end{array}%
\right]\!\!\!\!\!\!&\!\!\!\!\!\!\!=\!\!\!\!\!&\!\!\!\!\!\!\left[%
\begin{array}{c}
L^{(k,k)}[2]-L^{(k,k)}[1]\\
L^{(k,k)}[3]-L^{(k,k)}[1]\\
\vdots \\
L^{(k,k)}[K]-L^{(k,k)}[1]\\
\end{array}%
\right] \nonumber \\
&\!\!\!\!\!\!\!=\!\!\!\!\!&\!\!\!\!\!\!\left[\!\!\!%
\begin{array}{c}
{\bf h}^{(k)T}[2]{\bf V}^{(k)}[2]-{\bf h}^{(k)}[1]\\
{\bf h}^{(k)T}[3]{\bf V}^{(k)}[3]-{\bf h}^{(k)}[1]\\
\vdots \\
{\bf h}^{(k)T}[K]{\bf V}^{(k)}[K]-{\bf h}^{(k)}[1]\\
\end{array}%
\!\!\!\right] {\bf s}^{(k)} \nonumber \\
&=&\!\!\!\!{\bf H}^{(k)}_{eff}{\bf s}^{(k)},
\end{eqnarray}
where $k\in\{1,2,\ldots,K\}$. Recall that all beamforming matrices ${\bf V}^{(k)}[n]$ are independently generated with respect to ${\bf h}^{(k)T}[n]$ for $n\in\{2,3\ldots,K\}$, and we assumed that all elements of channel vectors are drawn from a continuous random distribution. From these facts, it is possible to show all elements of the effective channel for user $k$ at time slot $n$, ${\bf h}^{(k)T}[n]{\bf V}^{(k)}[n]$, are statistically independent. Further, since ${\bf h}^{(k)T}[n]{\bf V}^{(k)}[n]$ and ${\bf h}^{(k)}[1]$ are linear independent for $\forall n$ and $\forall k$, $\textrm{rank}\left({\bf H}^{(k)}_{\textrm{eff}}\right)=K-1$ with probability one. Therefore, by using a ZF decoder, user $k$ obtains ${\bf s}^{(k)}$. As a result, if the transmitter has one outdated and $K-1$ current CSI for the $K$-user $(K-1)\times 1$ MISO broadcast channel, the transmit delivers $K(K-1)$ data symbols to all users over $K$ time slots, which leads to achieve $K-1$ DoF in the system.

Using Theorem 4 and the same resource counting argument described in Section \ref{sec:tradeoff}, we establish the following proposition.

\textbf{Proposition 1}: The optimal $K-1$ of DoF are achieved for the $K$-user $(K-1)\times 1$ MISO broadcast channel if CSI feedback delay is less than $\frac{1}{K}$ of channel coherence time.

\begin{figure}
\centering
\includegraphics[width=3.4 in]{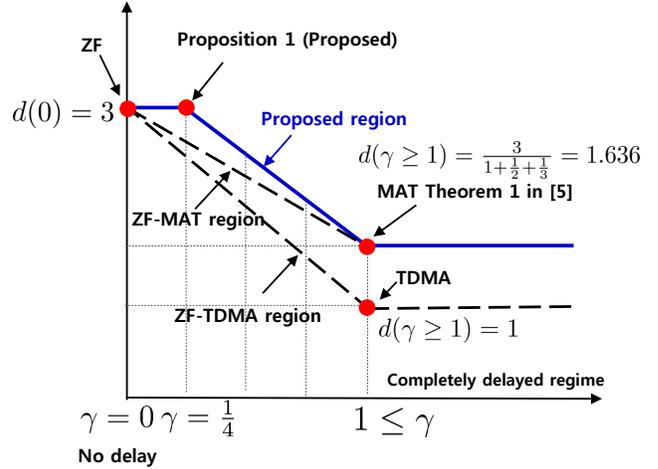}
\caption{CSI feedback delay-DoF gain trade-off for the 4-user
$3\times 1$ MISO broadcast channel.} \label{fig:Trade-off}
\end{figure}

For instance if $K=4$, the 3 of optimal DoF gain are obtained as long as CSI feedback delay is less than $20\%$ of channel coherence time as depicted in Fig. \ref{fig:Trade-off}.

\section{Conclusion} \label{sec:Conclusions}
We proposed a new algorithm that exploits both the current and outdated CSI for the MISO broadcast channel under a block fading assumption. We showed that the efficient exploitation of not only current CSI but also outdated CSI achieves the optimal DoF gain when the CSI feedback delay is less than a certain fraction of the channel coherence time. Using our results and leveraging results in \cite{Maddah-Ali2}, we proposed a CSI feedback delay-DoF gain trade-off for the 3-user MISO broadcast channel to provide an insights into the interplay between CSI feedback delay and system performance from a DoF gain perspective. From the derived trade-off result, we verified the intuition that a small CSI feedback delay should be negligeable.


\begin{thebibliography}{1}



\bibitem{Spencer}
Q. H. Spencer, A. L. Swindlehurst, and M. Haardt,
\newblock ``Zero-Forcing Methods for Downlink Spatial Multiplexing in
Multiuser MIMO Channels,''
\newblock {\em IEEE Tran. Signal Processing,} vol. 52,
no. 2, pp. 461-471, Feb. 2004.

\bibitem{Caire}
G. Caire and S. Shamai,
\newblock ``On the Achievable Throughput of a
Multiantenna Gaussian Broadcast Channel,''
\newblock {\em IEEE Trans. Inform. Theory,} vol. 49, no. 7, pp. 1691-1706, July
2003.


\bibitem{Jindal}
N. Jindal,
\newblock ``MIMO Broadcast Channels with Finite-Rate Beedback,''
\newblock {\em IEEE Trans. Inform. Theory,} vol. 52, no. 11, pp. 5045-5060, Nov. 2006.

\bibitem{Yoo}
T. Yoo, N. Jindal and A. Goldsmith,
\newblock ``Multi-Antenna Broadcast Channels
with Limited Feedback and User Selection,''
\newblock {\em IEEE Journal Selected Areas
in Communications,} vol. 25, pp. 1478-1491, Sep. 2007.


\bibitem{Maddah-Ali2}
M. A. Maddah-Ali and D. Tse,
\newblock ``Completely Stale Transmitter
Channel State Information is Still Very Useful,''
\newblock {\em Submitted to IEEE Trans. Inform. Theory,}  [Online]: arxiv:1010.1499v2.

\bibitem{Maleki}
H. Maleki, S. A. Jafar, and S. Shamai,
\newblock ``Retrospective Interference Alignment over Interference Networks,''
\newblock {\em IEEE Journal of
Selected Topics in Signal Processing}, vol. 6, no. 3, pp.228-240, June 2012.


\bibitem{Ghasemi}
A. Ghasemi, A. S. Motahari, and A. K. Khandani,
\newblock ``Interference
Alignment for the MIMO Interference Channel with Delayed Local
CSIT,''\newblock {\em Submitted to IEEE Trans. Inform. Theory}, Feb. 2011.
[Online]:arXiv:1102.5673.




\bibitem{Yang}
S. Yang, M. Kobayashi, D. Gesbert and X. Yi,
\newblock ``On the Degrees of Freedom of Time Correlated MISO Broadcast Channel with Delayed CSIT,''
\newblock {\em Submitted to IEEE Trans. Inform. Theory,} Mar. 2012. [Online]:arxiv.org/abs/1202.1909.

\bibitem{Gou_outdated}
T. Gou and S. Jafar,
\newblock ``Optimal Use of Current and Outdated Channel State
Information - Degrees of Freedom of the MISO BC with Mixed CSIT,''
\newblock {\em Submitted to IEEE Communications Letters,} Mar. 2012. [Online]:arXiv:1203.1301v1.

%
%
%
%




\bibitem{Birk}
Y. Birk and T. Kol,
\newblock ``Coding on Demand by an Informed Source (ISCOD) for Efficient Broadcast of Different Supplemental Data to Caching Clients,''
\newblock {\em IEEE
Trans. on Information Theory},vol. 52, no. 6, pp. 2825-2830, June 2006.

\bibitem{Yossef}
Z. Bar-Yossef, Y. Birk, T. S. Jayram, and T. Kol,
\newblock ``Index Coding With Side Information,''
\newblock {\em IEEE Trans. on Information Theory},vol. 57, no. 3, pp. 1479-1494, March 2011.

\bibitem{Rouayheb}
S. El Rouayheb, A. Sprintson, and C. N. Georghiades,
\newblock `'On the Index Coding Problem and Its Relation to Network Coding and Matroids,''
\newblock {\em Submitted to IEEE Trans. on Information Theory}, 2009.

\bibitem{Maleki2}
H. Maleki, V. R. Cadambe, S. A. Jafar
\newblock `'Index Coding: An Interference Alignment Perspective ,''
\newblock {\em Submitted to IEEE Trans. on Information Theory}, [Online]:arXiv:1205.1483v1.

\bibitem{Namyoon}
N. Lee and R. W. Heath Jr.,
\newblock `` Space-Time Interference Alignment and Degrees of Freedom
for the MIMO Broadcast Channel with Periodic CSI Feedback,''
\newblock {\em Submitted to IEEE Trans. Inform. Theory,} April 2012.

%
\end{thebibliography}
\end{document}